# Cross-sectional Stock Price Prediction using Deep Learning for Actual Investment Management

Masaya Abe
Nomura Asset Management Co,Ltd.
1-11-1 Nihonbashi, Chuo-ku, Tokyo,
103-8260, Japan
81 (0)3-4376-6049
masaya.abe.428@gmail.com

Kei Nakagawa
Nomura Asset Management Co,Ltd.
1-11-1 Nihonbashi, Chuo-ku, Tokyo,
103-8260, Japan
81 (0)3-4376-6049
kei.nak.0315@gmail.com

## ABSTRACT

Stock price prediction has been an important research theme both academically and practically. Various methods to predict stock prices have been studied until now. The feature that explains the stock price by a cross-section analysis is called a "factor" in the field of finance. Many empirical studies in finance have identified which stocks having features in the cross-section relatively increase and which decrease in terms of price. Recently, stock price prediction methods using machine learning, especially deep learning, have been proposed since the relationship between these factors and stock prices is complex and non-linear. However, there are no practical examples for actual investment management. In this paper, therefore, we present a cross-sectional *daily* stock price prediction framework using deep learning for actual investment management. For example, we build a portfolio with information available at the time of market closing and invest at the time of market opening the next day. We perform empirical analysis in the Japanese stock market and confirm the profitability of our framework.

## 1. INTRODUCTION

Stock price prediction has been an important research theme both academically and practically. Various methods to predict stock prices have been studied until now. These methods can be roughly divided into two aspects: time-series and cross-section analysis.

The first method analyzes past stock prices as time-series data and perform time-series analysis. The financial time-series analysis originally started from a linear model, such as the autoregressive (AR) model in which the parameters are uniquely determined [1]. As many nonlinear behaviors have been observed in actual financial time-series data, the generalized autoregressive conditional heteroscedasticity (GARCH [2]) model incorporating the time series structure into volatility has been used as one approach. In recent years, the GARCH model has been expanded to multivariate even for many stocks [3,4]. In addition, nonlinear models such as *k*-nearest neighbor [5], neural networks [6] and support vector machines [7] have been used for stock price predictions in terms of time-series analysis. These models not only strive to grasp economic implications academically but also strive to increase prediction accuracy practically. They especially try to grasp stock price fluctuation patterns by trial and error. These approaches have attracted attention for improving computing capabilities in recent years.

The second method performs cross-section (regression) analysis using cross-sectional data such as corporate attributes. The feature that explains the stock price by a cross-section analysis is called a "factor" in the field of finance. Many empirical studies in finance have identified which stocks having features in the cross-section relatively increase and which decrease in terms of price. The representative model that explains the cross-sectional stock prices is the Fama-French three-factor model [8,9]. They proposed that the cross-sectional structure of stock returns can be explained by three factors: beta (market portfolio), size (market capitalization), and value (price book-value ratio). Since then, many factors other than those in the Fama-French three-factor model were found one after another. As a result, [10] reported that over 300 factors were discovered until 2012. Moreover, most of these factors have been found in the last 10 years.

Although the factors that investors should consider are rapidly increasing, it is so difficult to simultaneously examine over 300 factors due to the curse of dimension. Besides, a linear regression model has been used in the financial field because of easy statistical handling and the robustness of the result. However, since the relationship between these factors and stock returns is complex [11], linear regression models have limited prediction accuracy. As non-parametric cross-sectional stock prediction studies [12-16], they used deep learning to combine various factors nonlinearly. They reported that the prediction accuracy and profitability can be improved by combining non-linearly using deep learning rather than simply combining various factors by linear regression.

However, these studies are limited in *monthly* stock price prediction and they are not in line with actual investment management. In this study, we present a cross-sectional *daily* stock price prediction framework using deep learning for actual investment management. And we perform empirical analysis in the Japanese stock market to confirm the effectiveness of our framework. In order to invest on a daily basis, we build a portfolio at a time when we can actually invest. For example, we build a portfolio with information available at the time of market closing and invest at the time of market opening the next day. In addition, the portfolio turnover rate is calculated and compared in order to consider the impact of transaction costs. A portfolio with a high turnover rate will have more transaction costs than with a portfolio with a lower rate.

The remainder of the paper is organized as follows. Section 2 summarizes related works. Section 3 provides a brief description of our prediction methodology. Section 4 shows the empirical study in the Japanese stock market. Section 5 concludes the paper.

## 2. RELATED WORK

Many studies on stock price prediction in terms of time-series analysis with machine learning have been published. For example, [17,18] showed that the shape of stock price fluctuation is an important feature in the prediction of future prices. They proposed a method to predict future stock prices with the past fluctuations similar to the current with indexing dynamic time warping method [19]. [20] created an automatic stock trading system in the Australian stock market. They used a neural network that decides when to buy or sell the stock. The inputs are four variables arising from the fundamental analysis: price-earnings ratio (PER), price book-value ratio (PBR), return on equity (ROE) and dividend payout ratio. The outputs are a strong signal that represents the expected returns of the predicted stock. [21] investigated how to predict stock indices by using support vector machines (SVMs) to learn the relationship among several technical indicators such as several moving averages and the stock index price. They used the grid search method to optimize the SVM model parameters. The experimental results show that transforming the input data space of SVM can bring good performance in finance engineering.

[22,23] presented a review of the application of several machine learning methods in finance. In their survey, most of these were forecasts in terms of time series analysis. However, there is no paper that deals with the prediction method in terms of a multi-factor model. There are many studies on daily stock price forecasting from the viewpoint of time series forecasting [24,25]. However, these studies are not actually investable because they trade at the closing price using information available after closing.

In terms of the cross-section analysis, [11] discussed the use of multilayer feedforward neural networks for predicting stock returns within the framework of the multi-factor model. [12,13] extended this model to deep learning and other machine learning model such as SVM and Random Forest. They investigated the performance of each machine learning method on the Japanese stock market. They showed that deep neural networks generally outperform shallow ones, and the best networks also outperform representative machine learning models. These works are only for use as a return model, and the problem is that the viewpoint of a risk model is lacking. [14] proposed the application of LRP [26] to decompose the attributes of the predicted return as a risk model. [15] extend this model to a time-varying multi-factor model with LSTM + LRP because they do not examine the influence on performance due to the approximation of LRP and not considering the time-dependency of factors. [16] proposed a deep transfer learning among multiple stock market regions. They showed that the deep transfer learning outperforms not only off-the-shelf machine learning methods but also the average return of major equity investment funds. However, these studies are limited in *monthly* stock price prediction and they are not in line with actual investment management. We implement a *daily* portfolio construction framework that invests at a time when we can actually invest and reduces the impact of rebalancing timing on performance.

## 3. DATASET AND METHODOLOGY

This section describes cross-sectional *daily* stock price prediction framework using deep learning for actual investment management.

### 3.1 Dataset

We prepare dataset for TOPIX500 Index constituents. The TOPIX500 Index comprises the large and mid-cap segments of the Japanese stock market. The index is also often used as an investment universe for overseas institutional investors investing in Japanese stocks. We use the 33 factors listed in Table 1.

**Table 1 List of Factors**

| No. | Factor |
|---|---|
| 1 | Return from previous day |
| 2 | Return from 2 days ago |
| 3 | Return from 3 days ago |
| 4 | Return from 5 days ago |
| 5 | Return from 10 days ago |
| 6 | Return from 20 days ago |
| 7 | Return from 40 days ago |
| 8 | Return from 60 days ago |
| 9 | Average trading value over the past 60 days |
| 10 | Average trading value over the past 5 days/60 days |
| 11 | Average trading value over the past 10 days/60 days |
| 12 | Average trading value over the past 20 days/60 days |
| 13 | Change in operating income forecast from 5 days ago |
| 14 | Change in operating income forecast from 10 days ago |
| 15 | Change in operating income forecast from 20 days ago |
| 16 | Change in target stock price forecast from 5 days ago |
| 17 | Change in target stock price forecast from 10 days ago |
| 18 | Change in target stock price forecast from 20 days ago |
| 19 | Book-value to Price Ratio |
| 20 | Earnings to Price Ratio |
| 21 | Dividend Yield |
| 22 | Sales to Price Ratio |
| 23 | Cashflow to Price Ratio |
| 24 | Return on Equity |
| 25 | Return on Asset |
| 26 | Return on Invested Capital |
| 27 | Accruals |
| 28 | Total Asset Turnover Rate |
| 29 | Current Ratio |
| 30 | Equity Ratio |
| 31 | Total Asset Growth Rate |
| 32 | Capital Expenditure Growth Rate |
| 33 | Investment to Asset |

These are used relatively often in practice. In calculating these factors, we acquire necessary data from Factset, WorldScope, Thomson Reuters, I/B/E/S. Forecast data is obtained from I/B/E/S to calculate No. 13-18. The actual financial data is acquired from WorldScope and Reuters Fundamentals (WorldScope priority). No. 19-33 are calculated on a monthly basis (at the end of month). The following are definitions of factors No.19-33.

No.19 = Net Assets/Market Value

No.20 = Net Profits/Market Value

No.21 = Dividends/Market Value

No.22 = Sales/Market Value

No.23 =Operating Cashflow/Market Value

No.24 =Net Profits/Net Assets

No.25 =Net Operating Profits/Total Assets

No.26=Net Operating Profits After Tax/(Debt + Net Assets)

No.27=-(Changes in Current Assets and Liabilities
    -Depreciation)/Total Assets

No.28=Sales/Total Assets

No.29=Current Assets/Current Liabilities

No.30=Net Assets/Total Assets

No.31=Change Rate of Total Assets from the previous period

No.32=Change Rate of Capital Expenditure
    from the previous period

No.33=Change Rate of Payments for acquisition of Tangible
    Fixed Assets from the previous period/Total Assets

## 3.2 Problem Formulation

To define the problem as a regression problem. For example, for stock $i$ in TOPIX500 Index constituents at day $t$ represented as $U_t$, 33 factors listed in Table 1 are defined by $x_{i,t} \in R^{33}$ as input values. The output value is defined by the next 5 day's stock return, $y_{i,t+5} \in R$. Note that $y_{i,t+5}$ is defined as $p^c_{i,t+5}/p^o_{i,t+1} - 1$ due to practical tradability. Here, $p^c_{i,t+5}$ denotes the closing price at day $t + 5$ and $p^o_{i,t+1}$ denotes the opening price at day $t + 1$. We define 5 days ahead stock return as output value to align with the portfolio construction method as describe later (Figure 1).

For data preprocessing, rescaling is performed so that each input value is maximally 1 (minimum≈0) by ranking each input value in ascending order by stock universe at each day and then dividing by the maximum rank value. Similar rescaling is done for output values $y_{i,t+5}$, to convert to the cross-sectional stock returns. We call the dataset ($x_{i,t}$,$y_{i,t+5}$) as one training data. Note that $x_{i,t}$ and $y_{i,t+5}$ are assumed to be the values after data preprocessing. This procedure is extended to using the latest $N = 1,000$ days rather than the most recent set of training data (one training set).

Our problem is to find a predictor $f$. We use the mean squared error (MSE) as the loss function and define $MSE_T$ when training the model at $T$ as follows:

$$MSE_T = \frac{1}{K} \sum_{t=T-N-4}^{T-5} \sum_{i \in U_t} \left( \left( y_{i,t+5} - f(x_{i,t}; \theta_T) \right) \right)^2 \quad (1)$$

$K$ is the number of all training data. $\theta_T$ is the parameter calculated by solving (1) and makes the form of a function $f$.

## 3.3 Prediction Models

We use deep learning as a model of the function $f$, and use ridge regression and random forest for comparison model. Details are as listed below.

**Deep Neural Network (DNN)**
DNN is implemented with an open source machine learning library TensorFlow [27]. For the hyperparameters, there are 6 patterns in total. For the hyperparameter, there are 6 patterns in total shown in Table 2. There are 3 patterns with hidden layer and 2 patterns with dropout rate and the number of epochs. We use the ReLU function [28] as the activation function, and Adam [29] for the optimization algorithm. Batch normalization [30] is applied to activation. The mini-batch size is set to 500. As for the starting point of the analysis, we initialize to generate the network weights from TensorFlow's function "tf.truncated_normal" set to mean 0 and standard deviation $\sqrt{2/M}$" ($M$ is the size of the previous layer).

**Table 2. The structure of DNN**

| Model | Hidden Layers (Dropout Rate) | Number of Epoch |
|---|---|---|
| DNN1 | 500-200-100-50-10 (50%-40%-30%-20%-10%) | 20 |
| DNN2 | 500-200-100-50-10 (50%-40%-30%-20%-10%) | 30 |
| DNN3 | 200-200-100-100-50 (50%-50%-30%-30%-10%) | 20 |
| DNN4 | 200-200-100-100-50 (50%-50%-30%-30%-10%) | 30 |
| DNN5 | 300-300-150-150-50 (50%-50%-30%-30%-10%) | 20 |
| DNN6 | 300-300-150-150-50 (50%-50%-30%-30%-10%) | 30 |

**Random Forest (RF)**
Random Forest is implemented with scikit-learn [31] with the class "sklearn.ensemble.RandomForestRegressor". For the hyper parameters, the number of features (max_features) is 11 (= 33/3), the number of trees(n_estimators) is 1,000 and the tree depth (max depth) is {3, 5, 7}. We denote RF1, RF2 and RF3 in order of increasing the tree depth.

**Ridge Rigression(RR)**
Ridge Regression is implemented with scikit-learn with the class "sklearn.linear_model.Ridge". For the hyper parameters, we set the regularization strength("alpha") to {0.1, 1, 10}. We denote RR 1, RR2 and RR3 in order of increasing the regularization strength.

We train the model by using the latest 1,000 sets of training data. To calculate the prediction, we substitute the latest input values into the model after training has occurred. The cross-sectional predictive stock return (score) of stock $i$ at day $T + 5$ is calculated from time $T$ by (2) substituting $x_{i,T}$ into the function $f$ in (2) with the parameter $\theta^*_T$, where $\theta^*_T$ is calculated from (1) with $N = 1,000$:

$$Score_{i,T+5} = f(x_{i,T}; \theta^*_T) \quad (2)$$

We construct investment portfolios with above scores.

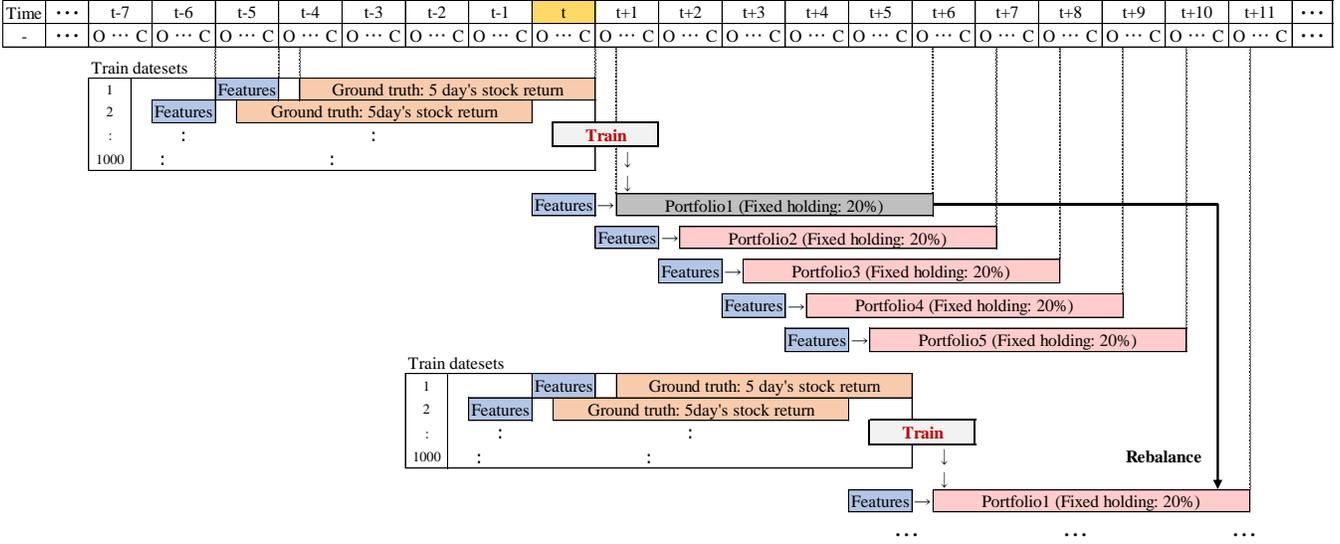

**Figure 1. Our Portfolio Construction Framework.**

## 3.4 Portfolio Construction Framework

In this paper, we consider two investment strategies that are widely used in the literature of finance [8-10]. Namely, (i) the long portfolio strategy, and (ii) the long-short portfolio strategy. We consider an equally-weighted portfolio, which is simple yet sometimes outperforms more sophisticated alternatives [32]. (i) The long portfolio strategy considered here buys the top quintile (i.e., one-fifth) scores of the stocks with equal weight aiming to outperform the average return of all the stocks. (ii) The long-short portfolio strategy not only buys the top quintile scores of the stocks but also sells the bottom quintile scores of the stocks. While the long-short portfolio cannot take advantage of the stock market growth, it is robust against a large market crisis (i.e., the financial crisis during 2007-2008) because of its market neutral position. Figure 1 shows our portfolio construction framework.

The performance of portfolios 1 to 5 with different rebalancing timings in Figure 1 varies depending on the daily stock market fluctuations. In order to reduce the chances of having no other good portfolio by holding only one of the five portfolios, we will hold all five portfolios equally. We rebalance one of five portfolios hold 20% every business day. The prediction models are updated every five business days.

## 3.5 Performance Measures

In evaluating the long portfolio strategy and the long-short portfolio strategy, we use the following measures that are widely used in the field of finance [33].

First, we define the return of long portfolio $R_t^L$ and long-short portfolio $R_t^{LS}$. Let $L_t \subset U_t: 1/5|U_t|$ be the long portfolio. The return from the long portfolio is defined as the average return of $L_t$.

$$R_t^L = 1/|L_t| \sum_{i \in L_t} y_{i,t}$$

Let $S_t \subset U_t: 1/5|S_t|$ be the short portfolio. The return from the short portfolio is defined as the average return of $S_t$

$$R_t^S = 1/|S_t| \sum_{i \in S_t} y_{i,t}$$

The return from the short portfolio is defined as

$$R_t^{LS} = R_t^L - R_t^S$$

Note that as shown in Figure 1, we calculate $R_t^L$ and $R_t^{LS}$ for each of the five portfolios and use their average as $R_t^L$ and $R_t^{LS}$ bellow.

Regarding the long portfolio strategy, the annualized return is the excess return (Alpha) against the average return of all stocks in the universe, the risk (tracking error; TE) is calculated as the standard deviation of Alpha and risk/return is Alpha/TE (information ratio; IR).

$$\text{Alpha} = \prod_{t=1}^{T}(1 + \alpha_t)^{\frac{250}{T}} - 1$$

$$\text{TE} = \sqrt{\frac{250}{T-1}(\alpha_t - \mu_\alpha)^2}$$

$$\text{IR} = \text{Alpha}/\text{TE}$$

Here, $\alpha_t = R_t^L - 1/|U_t| \sum_{i \in U_t} y_{i,t}$, $\mu_\alpha = 1/T \sum_{t=1}^{T} \alpha_t$.

Likewise, we evaluate the long-short portfolio strategy by its annualized return (AR), risk as the standard deviation of return (RISK), risk/return (R/R) as return divided by risk as for the long portfolio strategy.

$$\text{AR} = \prod_{t=1}^{T}(1 + R_t^{LS})^{\frac{250}{T}} - 1$$

$$\text{RISK} = \sqrt{\frac{250}{T-1}(R_t^{LS} - \mu_{LS})^2}$$

$$\text{R/R} = \text{AR}/\text{RISK}$$

Here, $\mu_{LS} = 1/T \sum_{t=1}^{T} R_t^{LS}$.

In summary, the return of the long (resp. long-short) portfolio is evaluated by Alpha (resp. AR), whereas the risk of the long (resp. long-short) portfolio is evaluated by TE (resp. RISK). We use the risk-normalized return (i.e. IR for the long and R/R for the long-short) that gives more reliable measure than the return itself.

We also evaluate maximum drawdown (MaxDD), which is yet another widely used risk measures [34], for both of the long portfolio strategy and the long-short portfolio strategy: Namely, MaxDD is defined as the largest drop from an extremum:

$$MaxDD = \min_{k \in [1,T]} \left( 0, \frac{W_k^{Port}}{\max_{j \in [1,k]} W_j^{Port}} - 1 \right)$$

$$W_k^{Port} = \prod_{i=1}^{k} (1 + R_i^{Port})$$

where $R_i^{Port} = \alpha_t$ (resp. $R_i^{Port} = R_t^{LS}$) for the long (resp. long-short) strategy.

For evaluating the rebalance amount, we calculate the one-way portfolio turnover (TN), which define as the average percentage of stocks traded in each period. TN from the long portfolio defines as

$$TN^L = \frac{1}{2(T-1)} \sum_{t=1}^{T-1} \sum_{i \in L_t \cup L_{t+5}} \left\| w_{i,t+5}^L - w_{i,t}^{L+} \right\|_1$$

where $w_{i,t+5}^L \in L_{t+5}$ is the portfolio weight at $t+5$ and $w_{i,t}^{L+} \in L_t$ is the long portfolio weight after considering stock price fluctuation between $t$ and $t+5$.

Likewise, as for TN from the long-short portfolio, we define $TN^{LS}$ as

$$TN^{LS} = TN^L + TN^S$$

$$TN^S = \frac{1}{2(T-1)} \sum_{t=1}^{T-1} \sum_{i \in S_t \cup S_{t+5}} \left\| w_{i,t+5}^S - w_{i,t}^{S+} \right\|_1$$

where $w_{i,t+5}^S \in S_{t+5}$ is the portfolio weight at $t+5$ and $w_{i,t}^{S+} \in S_t$ is the short portfolio weight after considering stock price fluctuation between $t$ and $t+5$. Finally, we average all five portfolio TN.

These performance measures are calculated daily on the basis of the opening during the prediction period from 4th January 2013 to 4th January 2018.

## 4. EMPIRICAL STUDY
### 4.1 Result of Long Portfolio
Table 3 shows the results of the long portfolio strategies. The bold letters represent the best of each method, and the best value in each column is also underlined.

The performance within DNN models is more variable than RF and RR models. This is because the high degree of freedom in the construction of DNN architecture with the large number of hyper-parameters. Compared with the number of epochs, the patterns of epoch 20 (DNN1, DNN3, DNN5) outperform epoch 30 (DNN2, DNN4, DNN6) in terms of Alpha and IR. These results show that the models trained until the number of epochs reach 30 tend to be overfitting. The difference in performance within RF models is smaller, and RR models are almost same results because the number of hyper-parameters to be adjusted is smaller than DNN models.

The best Alpha comes from RR2, and RR models outperform DNN and RF models. The results from IR are similar to Alpha but the best IR comes from DNN5. DNN models have lower TE and MaxDD, which indicate DNN models have an advantage in case of risk-averse strategies. Overall, RF models have lower TN and DNN models have higher TN.

**Table 3. Performance Summary of Long Portfolio**

| Model | Alpha | TE | IR | MaxDD | TN |
|---|---|---|---|---|---|
| DNN1 | 4.24% | 3.06% | 1.39 | **-3.15%** | 57.25% |
| DNN2 | 3.36% | 3.07% | 1.09 | -4.01% | 58.19% |
| DNN3 | **5.15%** | 3.53% | 1.46 | -3.84% | **55.23%** |
| DNN4 | 4.40% | 3.21% | 1.37 | -6.08% | 56.40% |
| DNN5 | 5.11% | 3.01% | **1.70** | -3.46% | 56.06% |
| DNN6 | 4.00% | **2.86%** | 1.40 | -4.14% | 56.54% |
| RF1 | **3.02%** | 4.47% | **0.68** | -8.23% | **44.31%** |
| RF2 | 3.02% | 4.49% | 0.67 | -8.75% | 46.01% |
| RF3 | 2.68% | 4.46% | 0.60 | -9.72% | 47.54% |
| RR1 | 5.95% | 3.82% | 1.56 | -4.03% | 49.19% |
| RR2 | **5.96%** | **3.82%** | **1.56** | **-4.03%** | **49.18%** |
| RR3 | 5.95% | 3.82% | 1.56 | -4.04% | 49.19% |

Figure 2 shows the daily cumulative aof the best IR portfolio within DNN, RF and RR. The red line (DNN5) is more stable throughout the period than the blue line (RR2) and orange line (RF1).

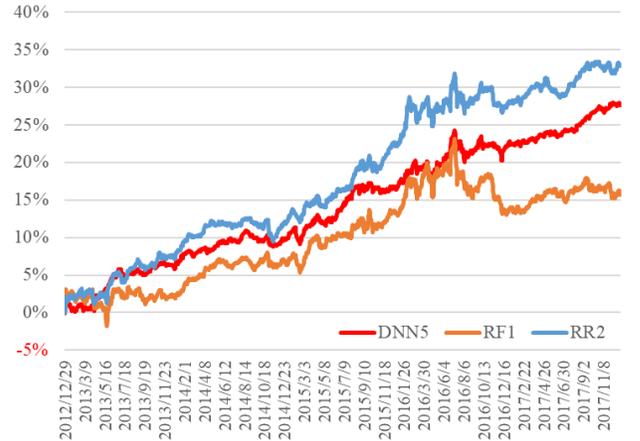

**Figure 2. The Daily Cumulative Alpha of the Best IR Long Portfolio within DNN, RF and RR.**

### 4.2 Result of Long-Short Portfolio
Table 4 shows the results of long-short portfolio strategies.

The difference in performance within each machine learning model tends to be similar to the results of long portfolio strategy; the performance of DNN models is more variable than RF and RR models.

The values of AR, RISK, and TN in all models are higher than long portfolio strategies because of taking more risk with adding a short-selling portfolio to a long portfolio. The values of MaxDD are getting worse for the same reason. The results from R/R in DNN and RF models are better than long portfolio strategies while RR models are worse. These results indicate the patterns of attractive stocks are different between long side and short side, therefore it is considered that DNN and RF models, which can take into account nonlinearity, can also earn profits on the short side.

The best Alpha and IR come from DNN5, and some of DNN models outperform RR models. The values of RISK and MaxDD in all DNN models are lower than RF and RR models. The result shows that DNN models are excellent in terms of low risk. Overall, RF models have lower TN and DNN models have higher TN, which are shown in long portfolio strategies. These results are consistent with previous researches [12-16] on *monthly* cross-sectional stock price prediction in the with deep learning.

**Table 4. Performance Summary of Long-Short Portfolio**

| Model | AR | RISK | R/R | MaxDD | TN |
|---|---|---|---|---|---|
| DNN1 | 8.74% | 6.04% | 1.45 | -7.00% | 111.39% |
| DNN2 | 6.83% | **5.56%** | 1.23 | **-6.01%** | 114.89% |
| DNN3 | 11.37% | 7.23% | 1.57 | -10.34% | **105.04%** |
| DNN4 | 10.00% | 6.63% | 1.51 | -10.40% | 108.54% |
| DNN5 | **11.42%** | 6.46% | **1.77** | -8.42% | 108.07% |
| DNN6 | 8.97% | 6.13% | 1.46 | -8.26% | 110.98% |
| RF1 | **7.28%** | 9.38% | 0.78 | -20.86% | **87.37%** |
| RF2 | 6.69% | 9.28% | 0.72 | -19.56% | 92.22% |
| RF3 | 6.86% | **9.23%** | 0.74 | **-17.88%** | 94.79% |
| RR1 | 10.98% | 8.50% | 1.29 | -13.37% | 95.36% |
| RR2 | **10.98%** | **8.50%** | **1.29** | **-13.36%** | **95.36%** |
| RR3 | 10.96% | 8.50% | 1.29 | -13.36% | 95.37% |

Figure 3 shows the daily cumulative return of the best R/R portfolio within DNN, RF and RR. The blue line (RR2) and orange line (RF1) fluctuate in return levels. Especially in the second half of the period, the orange line (RF1) is fallen significantly. On the other hand, the red line (DNN5) is a stable upward throughout the period.

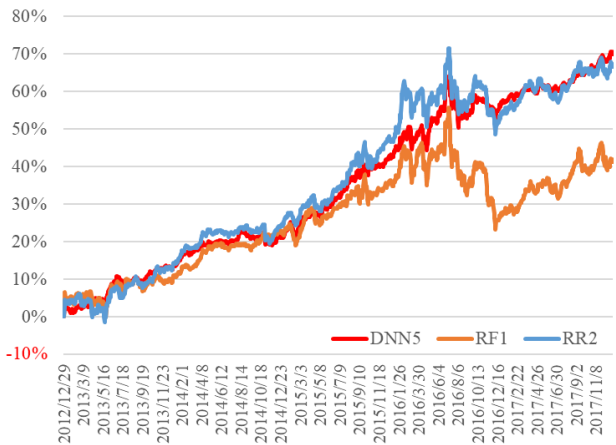

**Figure 3. The Daily Cumulative Return of the Best R/R Long-Short Portfolio within DNN, RF and RR.**

## 5. CONCLUSION

In this paper, we implement a cross-sectional *daily* stock price prediction framework using deep learning for actual investment management. We implemented a framework to predict five days ahead stock prices and build five portfolios rebalancing daily. The feature of our method is investable portfolio with information available at the time of market closing.

Our conclusions are as follows:

-The stock price prediction based on deep learning (DNN) has a larger performance variation due to the number of parameters than random forest (RF) and ridge regression (RR).

-DNN models have low TE and RISK. Especially, DNNs have mostly better R/R than RF and RR models.

-DNN models are higher turnover ratio than RF and RR models.

For further study, we examine how the performance changes in stock prediction period other than 5 days.